\documentclass[aps,pre,onecolumn,superscriptaddress]{revtex4}

\usepackage{bm}
\usepackage{mathrsfs}
\usepackage{graphicx}
\usepackage{color}
\usepackage{amsmath}
\usepackage{amssymb}

\begin{document}
\title{Universal proximity effect in target search kinetics in the
  few-encounter limit}

\author{Alja\v{z} Godec}
\email{agodec@uni-potsdam.de}
\affiliation{Institute of Physics \& Astronomy, University of Potsdam, 14476
Potsdam-Golm, Germany}
\affiliation{Department of Molecular Modeling, National Institute of Chemistry, 1000 Ljubljana, Slovenia}
\author{Ralf Metzler}
\email{rmetzler@uni-potsdam.de}
\affiliation{Institute of Physics \& Astronomy, University of Potsdam, 14476
Potsdam-Golm, Germany}

\begin{abstract}
When does a diffusing particle reach its target for the first time? This
first-passage time (FPT) problem is central to the kinetics of molecular
reactions in chemistry and molecular biology. Here we explain the behavior
of smooth FPT densities, for which all moments are finite, and demonstrate
universal yet generally non-Poissonian long-time asymptotics for a broad
variety of transport processes. While Poisson-like asymptotics arise
generically in the presence of an effective repulsion in the immediate
vicinity of the target, a time-scale separation between direct and
reflected indirect trajectories gives rise to a universal \emph{proximity
effect}: Direct paths, heading more or less straight from the point of release
to the target, become typical and focused, with a narrow spread of the
corresponding first passage times. Conversely, statistically dominant indirect
paths exploring the system size tend to be massively dissimilar. The initial
distance to the target particularly impacts gene regulatory or competitive
stochastic  processes, for which often few binding events determine the
regulatory outcome. The proximity effect is independent of details of the
transport, highlighting the robust character of the FPT features uncovered
here.
\end{abstract}

\maketitle 

\date{\today}   

\section{Introduction}

\noindent Exactly 100 years ago Marian Smoluchowski derived the rate
for the reactive encounter of two diffusing particles \cite{Smoluch}. Based
on Smoluchowski's concepts first passage time statistics characterize
the diffusion limitation of molecular
reactions \cite{Haenggi,Redner,Shles,Ralf,Kopelman,Szabo,oNChem,Schuss,benNaim,
Gleb,OlivierPRE,OlivierRep}. A case of particular relevance is cellular signaling
by specific molecules
\cite{OlivierRep,Holcman,otto,Mirny,Holcman2,GodSciRep,GodecPRE,Kurzinsky,Kolesov}.
The existence and impact of significant sample-to-sample fluctuations
in the precise timing of cellular
regulation, often at low-copy numbers of the signaling molecules, are by now well
established both
experimentally and theoretically
\cite{oNChem,Kurzinsky,Bialek,GodSciRep,GodecPRE,Stoch,Xie,Carlos2,GodMetzRC}. Remarkably,
despite the highly heterogeneous character of the motion of signaling molecules
in the nucleus and cell cytoplasm \cite{Gratton,Gorski,Franosch} the
overall signaling
kinetics appears to be universally stable, thus allowing cellular operation at a
remarkable precision \cite{BialekBK}. In the particular case of transcription
regulation the experimentally observed strong positional correlations between
pairs of downstream coregulated genes in both prokaryotic \cite{Kolesov} and eukaryotic
cells \cite{Fraser} indicate that proximity may indeed represent a tuning
mechanism, which was also supported theoretically in a 3-dimensional
setting \cite{otto}. According to \cite{oNChem} such proximity effects should only be
relevant for so-called \emph{geometry-controlled kinetics\/} when the molecules
explore their surrounding space in a \emph{compact\/} manner, that is, for recurrent
motion \cite{Liou} such as 1-dimensional diffusion or diffusion on
fractals \cite{oNChem}.
\begin{figure}
\begin{center}
\includegraphics[width=14.cm]{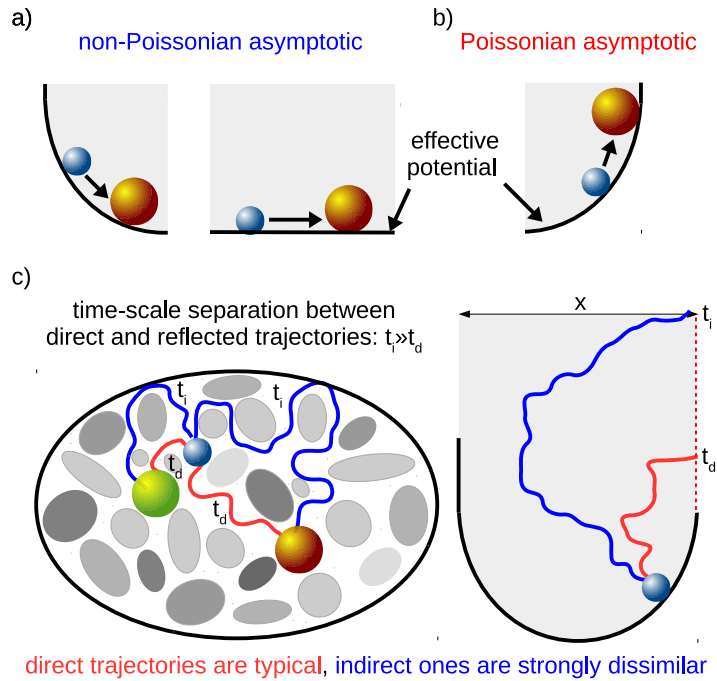}
\caption{Schematic of the two limiting forms of long-time FPT behavior
a) and b) depending on the effective potential experienced by the searcher in the vicinity of the
absorbing target (originating either from an
external force or from the geometric spurious drift in the radial direction). c) Schematics of two molecules (large spheres)
searching for a target (small sphere) with two different realizations of direct
and indirect trajectories $t_d$ and $t_i$, respectively, depending on
the initial distance to the target (see text for details). As soon as a time-scale separation
exists between $t_d$ and $t_i$ a robust proximity effect emerges, with
any two
$t_d$ being typical and very similar while any two $t_i$ tend to be
strongly dissimilar. This focusing of $t_d$ in turn enables faster
and more precise gene regulation.}
\label{scheme}
\end{center}
\end{figure}
Currently, there is no consensus on whether the eukaryotic chromosome
has a fractal structure \cite{Chro1,ChromRev}. In most bacteria the DNA is
segregated and spatially highly organized yet lacks a fractal organization
\cite{Toro}. In this context the immediate question arises whether
compact exploration is indeed necessary for the observed
regulatory precision \cite{Zlat,BialekBK}: namely, are proximity effects truly
limited to compact exploration---or could there in fact be a more general
mechanism to boost signaling speed and precision? 

Beyond the scope of transcription regulation, initial
  distances in spatial search processes are important in a large
  variety of contexts, ranging e.g. from locating enemy vessels in the
  ocean \cite{intermit} and search strategies of animals foraging for food \cite{foraging} to
the global spreading of diseases \cite{Lloyd}. Here, the initial
distance to the target is often essential \cite{Stanley,elite,intermit}, in particular, it plays an important
role in the foraging behavior of animals when food resources are sparse
\cite{foraging}, or in the predation behavior of marine vertebrates, including basking sharks \cite{sharks}, jellyfish and leatherback turtles \cite{turtles}, and southern
elephant seals \cite{seals}. Proximity becomes essential in the
first-come-first-serve sense corresponding to a destructive
search-limit \cite{Stanley}, i.e. when the target (food) disappears upon
the first time location by a searcher.

A key experimental factor in gene regulation is the
  fact, that in \emph{each realization} only the
  fastest few of the $10^2$ to $10^4$ transcription factor
  molecules relaying the biochemical signal to a gene determine the
  outcome (i.e. the cell's response) \cite{occup}. Similarly, the
  arrival of the fastest sperm cell at the egg cell decides the future of the new organism
  to be formed in each particular case. As we will demonstrate below, this situation is fundamentally very different from
  the question of how long it takes for the first signal to arrive \emph{on
  average} in an ensemble of independent non-destructive search realizations (see,
  e.g. \cite{oNChem,SidMeer,Katja}). This average scenario could correspond, e.g.
to ensemble experiments of molecular signaling effects in a colony of genetically
identical cells.
The characteristic arrival time of the fastest
  searcher in an ensemble of non-destructive search realizations is
  influenced to a dominant extent by massively dissimilar indirect
  trajectories, i.e. those
  which interact with the confining boundary before reaching
  the target \cite{oNChem,GodSciRep,Carlos2}. The statistics of
  such indirect trajectories was found to
  exhibit a qualitatively different behavior with respect to the
  compactness of spatial exploration in the limit of a large
  system size, and were therefore suggested to define
  two different universality classes \cite{oNChem}. How is this finding
  compatible with the fact that the FPT statistics in both cases follow an exponential asymptotic
  decay? And moreover, is the long-time behavior of the FPT statistics
  also sufficient to describe the first arrival kinetics in each given
  realization, such as encountered in transcription regulation?

Here we address these questions in two steps. First, without specifying
the actual motion pattern, we prove that the long-time behavior of all molecular target search processes in a confined and in
particular small domain belongs to the same
universality class, irrespective of whether the dynamics is compact
or not. Second, we analyze various specific types of motion
patterns and show how Poisson-like asymptotics arise if the particle experiences an
effective repulsion in the immediate vicinity of the target
(Fig. \ref{scheme} a and b). In contrast, the proximity effects we
uncover and quantify here naturally lead to an
accumulation of probability for observing direct trajectories for both
compact and non-compact Poisson-like dynamics (Fig. \ref{scheme}c), which enhances the speed and
precision of the target search process. In this sense,
to rationalize the proximity effect in gene regulation one needs to consider explicitly
the statistics of direct trajectories, i.e. those
  which do not interact with the confining boundary before reaching
  the target. Using experimentally relevant
parameters for transcription regulation we argue
that proximity effects may indeed represent a universal means for tuning
cellular signaling kinetics.    

The paper is structured as follows. First we represent a general
theory for the long-time behavior of smooth densities, for which all
moments are finite, and establish a unification of the previously
proposed universality classes in the finite volume case.
Next we demonstrate, on the hand of
exact result for a variety of different passive and driven diffusion
processes, the existence of a universal proximity effect in the
presence of a time-scale separation between direct and indirect
trajectories. Next, we discuss the biophysical implications of these
results in the context of transcription
regulation. We conclude with a summary of our results and offer a perspective on the
importance of our results for future studies.

\section{Theory and General Results for the Long-time Behavior}

\noindent To address the problem in the most general setting we impose only
mild constraints, which are warranted by the
physical setting: we assume the finiteness of the moments $\langle
T^n\rangle$ of the FPT
density. This means that the dynamics has a
finite natural time-scale, as expected in a finite volume.
Then the Laplace transform of the FPT
density, $\tilde{\wp}(s)=\hat{\mathcal{L}}[\wp:t\to s]$, of a particle starting at position $x_0$
to arrive at the target at $x_a$ (Fig.~S1), admits a moment expansion
$\tilde{\wp}(s)=\sum_{i=0}^{\infty}\langle T^n
\rangle(-s)^n/n!$. Moreover, it  has the following representation in the form of a quotient of convergent power
series
\begin{equation}
\tilde{\wp}(s)=\frac{g(s;x_0)}{h(s;x_a)}\equiv\frac{\sum_{k=0}^{\infty}g^{(k)}(x_0)s^k/k!}{\sum_{k=0}^{\infty}h^{(k)}(x_a)s^k/k!}
\label{powser}
\end{equation}
(see S1 in Supplementary information, SI),
where the superscript $(k)$ denotes the $k$th derivative with
respect to $s$ evaluated at $s=0$ \cite{Feller}. Here, we
limit the discussion to hyper-spherically symmetric Markov systems but our
general result can be readily applied to any geometry given the
knowledge of $g^{(k)},h^{(k)}$. 
By Cauchy's theorem the
asymptotic behavior of $\wp(t)$ is determined by the root of $h(s)$
which is closest to the origin at $s=0$. 
        
In a confined domain the transport operator $\hat{L}$ for such dynamics has
a discrete, real and non-degenerate eigenvalue spectrum
$\{\lambda_i\}$ with eigenvectors $\psi_i$. Moreover, $\lambda_i<0$
for all $i$ due
to the absorbing boundary \cite{Gardiner}. The long-time behavior of $\wp(t)$
is determined in terms of the lowest eigenpair
$(\lambda_0,\psi_0)$: 
$\wp(t)\sim -\mathcal{K}\psi_0(x_0)\frac{\partial\psi_0(x_a)}{\partial
  x}\mathrm{e}^{-\lambda_0 t}$, where the constant $\mathcal{K}$ takes into account the specific properties of
the given system and $\sim$ denotes asymptotic equality.

By necessity $\lambda_0$ is a simple pole. 
To find it, we must invert $h(s)$ in the vicinity of $0$ for
negative real $s$. As long as $\tilde{\wp}(s)$ does not have a branch
cut along the negative real axis---an assumption, which is confirmed
below---this can be done exactly in the form of a 
Newton series (for details see SI-S2). The first step of this procedure consists of obtaining the solution $-\lambda_0$ of the
  non-linear equation $h(s;x_a)=0$. The exact result is given by the
Newton series (see SI-S2)
\begin{equation}
\label{exact}
\lambda_0(x_a)=\sum_{k=1}^{\infty}\frac{h^{(0)}(x_a)^k}{h^{(1)}(x_a)^{2k-1}}\frac{\mathrm{det}\mathcal{A}_k}{(k-1)!},
\end{equation}
where $\mathcal{A}_k$ is an almost triangular square matrix with
elements 
\begin{equation}
\label{element}
\mathcal{A}_n(i,j)=\frac{h^{(i-j+2)}\Theta(i-j+1)}{(i-j+2)!}\left[n(i-j+1)\Theta(j-2)+i\Theta(1-j)+j-1\right].
\end{equation}
Here $\Theta(n)$ is the discrete Heaviside step function and we use
the symbolic convention $\mathrm{det}\mathcal{A}_1\equiv 1$ (see also SI, Eq.~(S11)). The series (\ref{exact}) is
generally rapidly converging, e.g. for 1-dimensional Brownian motion it converges
to within $99.4$\%\ of the known result $\frac{\pi^2}{4}$ \cite{Redner,Carlos2} already at
$k=4$. 
As long as the particle is not strongly
biased towards the target, in which case $k \sim\mathcal{O}(100)$,
as $\wp(t)$ tends to a delta function, $\lesssim 10$ terms are
typically sufficient to achieve convergence to numerical
precision. Explicitly,
the lowest order terms of Eq.~(\ref{exact}) have the form
\begin{equation}
\label{first}
\lambda_0(x_a) \sim  \frac{h^{(0)}(x_a)}{h^{(1)}(x_a)}\left(1-\frac{h^{(0)}(x_a)}{2}\frac{h^{(2)}(x_a)}{h^{(1)}(x_a)^2}\left[1+\mathcal{\mathcal{O}}\right]\right),
\end{equation}
where the first terms of the remainder $\mathcal{O}$ read
\begin{equation}
\label{rem}
\mathcal{O} = \frac{h^{(0)}h^{(2)}}{h^{(1)2}}\left(1+\frac{5}{4}\frac{h^{(0)}h^{(2)}}{h^{(1)2}}\right)-\frac{11}{18}\frac{h^{(0)2}h^{(3)}}{h^{(1)3}}+\frac{1}{12}\frac{h^{(0)2}h^{(4)}}{h^{(2)}h^{(1)2}}+\ldots
\end{equation}\\
\indent After obtaining $\lambda_0$, we can isolate the leading order
term of $\tilde{\wp}(s)$ under the quite loose condition
$\lim_{n\to\infty}\frac{g^{(n)}}{h^{(n)}}<\infty$, which is fulfilled
in all the cases studied here and appears to follow generally from the
positivity and finiteness of $\langle T^n \rangle$ (for a detailed
discussion and justification
see SI-S3). Now it is essentially straightforward to 
invert the Laplace transform
using Cauchy's theorem to obtain the exact long-time behavior of the
FPT density (for a detailed proof see SI-S2). The result reads
\begin{equation}
\label{theorem}
\wp(t)%&\sim& \lim_{s\to
\sim \lim_{k\to\infty}\frac{\sum_{l=0}^{k-1}\frac{1}{l!}\left[g^{(l)}-h^{(l)}\frac{g^{(k)}}{h^{(k)}}\right](-\lambda_0)^l}
{\sum_{l=0}^{k-1}\sum_{m=1}^{k-l}\frac{1}{(l+m)!}h^{(l+m)}(-\lambda_0)^{l+m-1}}\mathrm{e}^{-\lambda_0t}.
\end{equation}

Note that the moments $\langle T^n \rangle$ can
be obtained recursively from the coefficients $g^{(k)}$ and $h^{(k)}$ and vice versa (see
Eq.~(S2) in  SI-S1 for details). Eqs.~(\ref{exact}) to (\ref{theorem})
are our first main result. These expressions are important as they enable us to obtain
the exact long-time behavior of any smooth density, which
has an exponential asymptotic decay.
%, from its corresponding Laplace
%transform or the moments $\langle T^n \rangle$. 
Eqs.~(\ref{exact}) and (\ref{theorem}) are \emph{universal} in the
sense that the obtained asymptotic exponential decay
holds for any first passage process with a smooth density, for which
all $\langle T^n\rangle < \infty$. Note that they also hold
irrespective of the initial distance to the target. Namely, while in general the Laplace transform of
some arbitrary asymptotically exponential density $\tilde{\wp}(s)$ may display
multi-scaling with respect to $s$ (see, e.g. \cite{Godec_LREs}), smoothness and finiteness
of $\langle T^n\rangle$ assure exactly the validity of the power series
in Eq.~(S1) at least in the domain $-\lambda_0 - \delta < \mathrm{Re}(s) < \lambda_0$ for some real $\delta>0$
\cite{Godec_LREs}. This satisfies the requirements for the above
result to apply. Our general
result should therefore also apply to
non-Markovian dynamics as soon as all moments of the FPT density are
finite. However, while results for the mean FPT for a broad class of Gaussian non-Markovian processes in
bounded domains were obtained recently in \cite{Olivier_Nature} and
it is also known that a specific class of confined Gaussian non-Markovian
processes indeed has exponential long-time asymptotics \cite{Sokolov}, the
properties of higher moments and the long-time
asymptotics of FPT density for a general  non-Markovian processes remain elusive.

\subsection{\emph{Unification of universality classes}}

\noindent To first demonstrate the universality implied by Eq.~(\ref{theorem}) we analyze
the FPT behavior in a broad variety of spatially confined systems, in
which the stochastic dynamics ranges from
compact to non-compact cases, and we also consider diffusion in external
fields. Specifically, we study unbiased diffusion in hyper-spherical
domains in the presence of a centered absorbing target in dimensions
$1\le d \le 3$. As an extension to Brownian motion we also examine
diffusion in disordered media and fractals with fractal dimension
$d_f$ \cite{oNChem,Procc,bAv}. Thus we analyze the scaling
limit of a Markovian random walk, whose mean squared displacement
scales as $\langle r^2 (t) \rangle \propto t^{2/d_w}$, where $d_w\ge 2$ denotes the random walk dimension and $d_w>2$
describes subdiffusive motion \cite{bAv}. The various different regimes of the FPT behavior are captured by the parameter $\nu=\frac{3}{2}-\frac{d_f+1}{d_w}$.
Among biased diffusion
processes we analyze $1$-dimensional diffusion in a linear potential
(Taylor dispersion) \cite{Redner}, $2$-dimensional diffusion in
a radial potential flow \cite{Redner,Koplik} and the Kramer's
escape of an overdamped particle
from a harmonic potential
\cite{Siegert,Kramers,Haenggi,Schulten,Eli,Grebenkov}. 

Particle-based computer simulations are carried out to corroborate the
analytical results. Passive diffusion and $2$-dimensional radially biased diffusion is
simulated in terms of the (squared) Bessel process with parameter
$\nu$ by numerically integrating the Ito stochastic 
equation $dY_t=(1-2\nu)/Y_t+\eta_t$ with zero mean Gaussian white
noise of variance $\langle\eta(t)\eta(t')\rangle=2D\delta(t-t')$ up
to the FPT $Y_t=r_a$ \cite{Yor}. Biased $1$-dimensional diffusion and
the Kramer's escape from a parabolic potential are
simulated by numerically integrating the Ito stochastic
equations: $dY_t=v+\eta_t$, and
$dY_t=-m\gamma^{-1}\omega^2Y_t+\eta'_t$,
respectively. $1$-$5\times 10^4$ realizations were taken to measure $\wp(t)$. 

\begin{figure*}
\begin{center}
\includegraphics[width=16.cm]{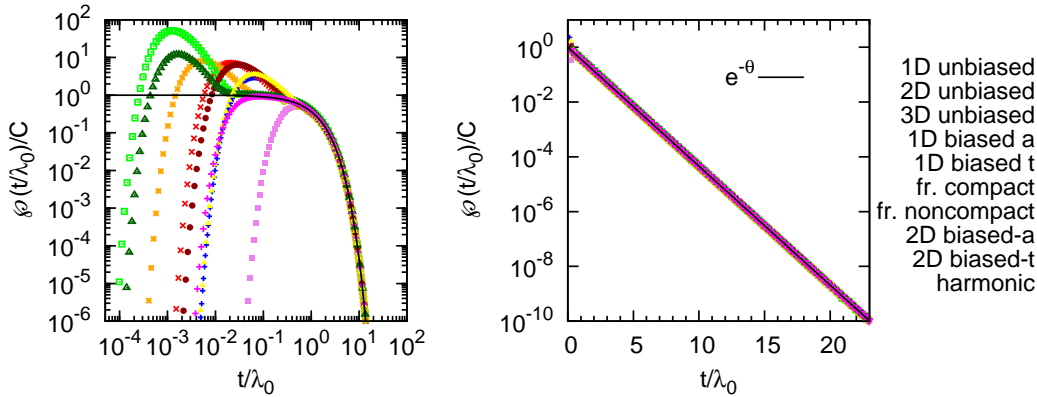}
\caption{Asymptotic collapse and short-time non-universality for the rescaled FPT densities $\wp(\theta=t/\lambda_0)/\mathcal{C}$ for the various models described
  in the text. The parameters  are $x_0=0.4$ (1D diffusion), $x_a=0.2,x_0=0.5$ (2D diffusion), $x_a=0.1,x_0=0.4$ (3D diffusion),
  $x_0=0.3,\mathrm{Pe}=3$ (1D biased away),
  $x_0=0.4,\mathrm{Pe}=-3$ (1D biased towards), $\nu=-0.25,x_a=0.2,x_0=0.45$   (non-compact diffusion of fractals), $\nu=0.25,x_a=0.2,x_0=0.45$
  (compact diffusion of fractals), $x_a=0.2,x_0=0.5,\mathrm{Pe}=3$   (2D radial potential flow away from the target), $x_a=0.2,x_0=0.5,\mathrm{Pe}=-1.8$
  (2D radial potential flow towards the target) and  $x_0=-1,x_a=2.8$ (overdamped escape from a harmonic potential).}
\label{collapse}
\end{center}
\end{figure*}

According
to Eq.~(\ref{theorem}) the FPT densities for
these fundamentally different processes should collapse, after
rescaling, to the asymptotic unit exponential $\wp(\theta=t/\lambda_0)/\mathcal{C}\sim \mathrm{e}^{-\theta}$ 
where $\mathcal{C}$ represents the prefactor in \eqref{theorem}.
In excellent agreement, Fig.~\ref{collapse} demonstrates
that the rescaled FPT densities indeed collapse to a master curve at
long times for all processes considered here.
Note the excellent agreement of the data with the predicted single-exponential
Therefore, the long-time behavior of all first passage processes with finite moments
unifies in a single universality class---the family of exponential
densities. How can this be reconciled with the two different
universality classes obtained in Ref.~\cite{oNChem}? The solution of this
seeming contradiction lies in the exact way the limits are taken, as
we detail now. 

In Ref.~\cite{oNChem} the authors considered passive Markovian diffusion in a
(generally fractal) medium with (fractal) dimension $d_f$ in a \emph{large volume
limit\/} of the confining domain. Since such Markovian diffusion is self-similar in
arbitrary dimension \cite{Varadhan}, the results are invariant with
respect to the exact volume (or equivalently, the radius of the domain $R$) if we express time
in natural units $\tau=R^{d_w}/D_{d_w}$,  where we defined the (generalized) diffusion
coefficient $D_{d_w}$. Therefore we can,
without any loss of generality, consider the problem in a domain of
unit radius, and express spatial coordinates in units of $R$ (see
below for details of the transformation). In this
invariant setting the large volume limit in Ref. \cite{oNChem}
corresponds to the limit when the dimensionless target size $x_a$ tends to
be very small.\\ 
\indent In a passive fractal medium the particle
experiences the aforementioned effective, $d_f$- and $d_w$- dependent geometric
outward radial bias (called the 'centrifugal drift' in
\cite{Redner}). As we will show below, the mathematical
origin for the different limiting behaviors with respect to $d_f$ and
$d_w$ found in \cite{oNChem} is actually rooted in the
fact, that the limit $x_a\to 0$ corresponds to a singular perturbation
in the case of non-compact exploration, but merely to a change
of the time unit for
compact exploration. Mathematically, the limit of a large volume
thus has a different meaning for compact and
non-compact exploration.

Another important point in this context follows from the exact limiting behavior of the coefficients
$h^{(k)}$ and the rate of convergence of the Newton series
(\ref{exact}) when the effective potential experienced by the searching particle
in the vicinity of the absorbing target (originating either from an
external force or from the geometric spurious drift in the radial direction) becomes sufficiently
repulsive, such that at least locally in the vicinity of the target the
motion is effectively non-recurrent. 
As we prove below, in this limit 
already the first correction $\frac{h^{(2)}(x_a)}{h^{(1)}(x_a)^2}$ to
the leading order behavior in
Eq.~(\ref{first}) converges uniformly to $0$. Similarly,
the prefactor in Eq.~(\ref{theorem}) converges with the first term, and  Eq.~(\ref{theorem}) reduces to 
\begin{equation}
\label{general}
\wp(t)\sim\frac{\langle T(x_0)\rangle}{\langle T(x^{\dagger})\rangle}\frac{1}{\langle T(x^{\dagger})\rangle}\mathrm{e}^{-t/\langle T(x^{\dagger})\rangle}.
\end{equation}
Here $\langle T(x_0)\rangle$ and $\langle T(x^{\dagger})\rangle$ are the mean FPT from the initial location $x_0$
and the location $x^{\dagger}$ of the potential minimum. If the potential is monotonic or when the bias
has a geometric origin then $x^{\dagger}$
corresponds to the location of the confining boundary.

In the limit $x_a\to 0$ this effective bias becomes
sufficiently large for $\frac{3}{2}<\frac{d_f+1}{d_w}\le 2$ (i.e. for
so-called non-compact spatial exploration \cite{oNChem}) and leads
to Eq.~(\ref{general})---the emergence of Poisson-like
asymptotics (for details, explicit results and proofs see below and
SI-S3). In contrast, as we show below in the case of compact
exploration, i.e., $0<\frac{d_f+1}{d_w}\le \frac{3}{2}$, the
coefficients $h^{(k)}$ become independent of $x_a$ in the limit
$x_a\to 0$ as the effective outward bias vanishes. As a result, higher
order terms in Eq.~(\ref{exact}) need to be considered as well.

Therefore, the long-time behavior of the FPT density for diffusion in generally
fractal media in fact falls into a single universality class, i.e. it can be
described by the single universal equation (\ref{theorem}). It is only the
behavior of the coefficients in Eqs.~(\ref{exact}) to (\ref{theorem}), which
in the limit of a vanishing target (or large volume, in the language
of \cite{oNChem}) depends on the compactness of the exploration of space. In fact one can check, by a suitable
identification of parameters and a corresponding change of units, that
our single universal result reproduces both qualitative regimes of the
limiting long-time behavior found in \cite{oNChem}. In that sense our results in Eqs.~(\ref{exact}) to (\ref{theorem}) fully corroborate and unify the findings in Ref.
\cite{oNChem}. In addition, we here present the exact forms of the short- and
intermediate-time FPT behavior.

Here we generalize the cases considered in \cite{oNChem} and demonstrate explicitly
that the single universal result also captures diffusion in external
fields. In addition, we prove the reduction of Eq.~(\ref{theorem}) to
the Poisson-like behavior (\ref{general}) for a sufficiently strong locally
repulsive potential. An important point we will make is that in all these
cases the long-time behavior is in fact insufficient to describe and
explain the \emph{kinetics in the few-encounter regime}, such as those
occurring in gene regulation, and that the cognizance of the full FPT
distribution $\wp(t)$ is needed to quantitatively understand the system.

\section{Time-scale separation and the proximity effect}

As we will show later on, the long-time behavior of $\wp(t)$ is not
sufficient to explain the proximity effect. Note that even the emergence of Poisson-like asymptotics in
Eq.~(\ref{general}) in general does not imply an existence of a single
time scale in the kinetics. In the presence of a time-scale separation
between direct and reflected indirect trajectories (Fig.~\ref{Classes}c)
proximity effects emerge for biased as well as for compact and non-compact
passive motion.

In contrast to the long-time asymptotic, the short-time behavior of
$\wp(t)$, involving the direct trajectories from the initial position towards the target which
do not feel the presence of the confinement, 
shows no such universality.
While it contains in all the studied cases
the L\'evy-Smirnov density
\begin{equation} 
\label{LSmirn}
\Phi(t;x_0-x_a)=\frac{|x_0-x_a|}{2\sqrt{\pi
    t^3}}\exp\left(-\frac{|x_0-x_a|^2}{4t}\right)
\end{equation} 
describing trajectories propagating
directly towards the target, in general this expression is multiplied by
a non-universal prefactor and potentially also a $t$-dependent function
(see below). The behavior in Eq.~(\ref{LSmirn}) is generally known as
the Sparre Andersen result and describes the universal behavior of the
free-space FPT
statistics of 1-dimensional Markov processes with symmetric steps
\cite{GodSciRep,Sparre,Koren} and thus even holds for L\'evy flights with a
diverging variance of the step length \cite{Koren}.

Depending on the initial distance to the target, there
also exists an intermediate power-law regime corresponding to
trajectories which make a brief excursion away from the target. The range of validity is strongly parameter
and system-dependent, but the resulting proximity effect is
universal. We now address the aforementioned specific systems
explicitly. The practical consequences of the proximity effect for
gene regulation are discussed in the last subsection.

\noindent \subsection{\emph{Passive diffusion}} We focus first on the unbiased
diffusion of a particle in a hyperspherically symmetric 
domain of radius $R$ and a perfectly absorbing target with radius $r_a$
in the center. In our discussion we also include the case of
subdiffusion and consider the possibility of a fractal medium with fractal dimension $d_f$ \cite{Procc,bAv,oNChem}. 
For this model the qualitative regimes of the FPT behavior are captured by the
single parameter $\nu=\frac{3}{2}-\frac{d_f+1}{d_w}$ (see
Eqs.~(\ref{sh_fract}) to (\ref{expo}) as well as SI-S3). If we limit
ourselves to embedding dimensions $1 \le d \le 3$ and recall that $d_w\ge 2$ and
$0\le d_f \le d$, the possible values of $\nu$ fall in the
interval spanned between $3$- and $1$-dimensional Brownian motion and hence
$-\frac{1}{2} \le \nu < \frac{3}{2}$. 
Moreover, the size $R$ of the confining domain factors out, as it merely
sets the characteristic diffusion time
$\tau=R^{d_w}/D_{d_w}$. Hence, without loss of generality we introduce dimensionless units
$x_i=\frac{r_i}{R}$ as well as the short-hand notation $\hat{x}=\frac{2}{d_w}x$ and express time in units of $\tau$. 
The short-time asymptotic reads (for a derivation see SI-S3) 
\begin{equation}
\label{sh_fract}
\wp(t)\sim
\left(\frac{x_0}{x_a}\right)^{\nu-1/2}\Phi(t;|\hat{x}_0-\hat{x}_a|),
\end{equation}
valid for $t\lesssim \hat{x}_a^2,\hat{x}_0^2$.
Apart from the $t-$independent prefactor this result exhibits the
Sparre Andersen form in Eq.~(\ref{LSmirn})
\cite{GodSciRep,Sparre}. On the level of a stochastic
  differential equation the effective
process underlying the mean field limit of diffusion in
hyperspherically symmetric fractal media is indeed a $1$-dimensional
Markov process in the radial coordinate (i.e. the squared Bessel process with parameter $\nu$
\cite{Yor}). Yet, it is not symmetric due to the spurious geometric outward
drift in the radial direction. This fact readily explains the additional
dimensionality-dependent prefactor.
Note that for
$\nu>\frac{1}{2}$ the prefactor tends to be large in the limit
$x_a\to 0$ and $x_0 \not\simeq x_a$. 

Following this short-time behavior, there is
an intermediate regime for
$[\Gamma(1+|\nu|)/|\Gamma(2-|\nu|)|]^{1/(1-|\nu|)}(\hat{x}_0/2)^2\ll
t\ll 1$ (SI-S3),
%\begin{eqnarray}
\begin{equation}
\label{int_fract}
\wp(t)\sim
\left(\frac{x_0}{x_a}\right)^{2\nu\Theta(-\nu)}\left[\left(\frac{\hat{x}_0}{2}\right)^{2|\nu|}-\left(\frac{\hat{x}_a}{2}\right)^{2|\nu|}\right]t^{-(1+\nu)}
\left\{\frac{1}{\Gamma(|\nu|)}-\frac{(\hat{x}_a/2)^{2|\nu|}}{\Gamma(-2|\nu|)}\left(\frac{\Gamma(1-|\nu|)}{\Gamma(1+|\nu|)}\right)^2t^{-\nu}\right\},
%\end{eqnarray}  
\end{equation}
for $\nu\ne 0$. $\Theta(z)$ and $\Gamma(z)$ denote the
Heaviside and Gamma functions. In
the limit $\nu=0$ (including 2-dimensional Brownian motion with $d_f=d_w=2$) we recover a logarithmic correction to the
power-law scaling, $\wp(t)\sim 2\log(x_0/x_a)/(t\log^2(t))$. Note that the leading
terms of Eqs.~(\ref{sh_fract}) and (\ref{int_fract}) coincide for
$|\nu|=1/2$ (including 1-dimensional Brownian motion with $d_f=1$ and $d_w=2$) in the regime $t\gg \hat{x}_0,\hat{x}_a$. Moreover,
Eqs.~(\ref{sh_fract}) and (\ref{int_fract}) together describe direct
FPT trajectories (Fig.~\ref{Classes}), which do not touch the outer
boundary of the system.

For any $\nu$ the long-time asymptotic corresponds exactly to the universal
result in Eq.~(\ref{theorem}) with the coefficients $g^{(k)}$ and $h^{(k)}$ given in
Eqs.~(S27)--(S28) in SI and agrees perfectly with the simulations results in
Fig.~\ref{collapse}. As mentioned above this generalizes and unifies the results in
Ref.~\cite{oNChem}.

To explain the different limiting behavior of $g^{(k)}$ and $h^{(k)}$ we focus
first on so-called non-compact dynamics, where the
particle sparsely explores the space \cite{oNChem} and hence $d_f>d_w$ or
$-\frac{1}{2}\le\nu <0$. Examples include normal diffusion in $d\ge2$
or diffusion and subdiffusion  on fractal objects when the fractal dimension is larger
than the walk dimension, $d_f>d_w$ \cite{oNChem}. Here the
aforementioned effective potential at the target site is the geometric or centrifugal bias away from the
center controlled by the target radius $x_a$. From the exact solutions for
 $g^{(k)}$ and $h^{(k)}$ (given in SI-S3)
it follows in the regime $d_f>d_w$, that in the limit $x_a\to 0$ the correction term in
Eqs.~(S32) and (S33) vanishes and we find for arbitrary $x_0$
\begin{equation}
\label{expo}
\lambda_0 \sim \frac{4|\nu|(1+|\nu|)}{\hat{x}_a^{-2|\nu|}-|\nu|-1},\quad
\mathcal{C}\sim \frac{4|\nu|(1+|\nu|)[\hat{x}_a^{-2|\nu|}-\hat{x}_0^{-2|\nu|}-|\nu| \hat{x}_0^2]}{(\hat{x}_a^{-2|\nu|}-|\nu|-1)^2}.
\end{equation}
As an important example we consider the limit $\nu\to 0$ (i.e. null-recurrent dynamics \cite{Liou}) and make use of the identity
$\lim_{\delta\to 0} (1-x^{\delta})/\delta=-\log(x)$. Thus we recover
smoothly the limiting case of critical compactness
($\nu=0$),
\begin{equation}
\label{expo2}
\lambda_0 \sim \frac{2}{-\log(\hat{x}_a)}, \quad
\mathcal{C}\sim
\frac{\log\left(\frac{x_0}{x_a}\right)-\hat{x}_0^2}{\log^2(\hat{x}_a)}.
\end{equation}
Eqs.~(\ref{expo}) and (\ref{expo2}) have the form of
Eq.~(\ref{general}) and the prefactor in Eqs.~(\ref{expo}) and (\ref{expo2}) tends to be
very small for $x_0 \simeq x_a$. For example, in the
  case of normal diffusion
  in 3 dimensions ($d_f=3$, $d_w=2$) we have $\langle
  T(x_0)\rangle=\frac{1}{3}(x_a^{-1}-x_0^{-1}+(x_a^2-x_0^2)/2)$
  \cite{GodSciRep}. Thus we
  find in the limit $x_a\to 0$ for $\nu=-\frac{1}{2}$ that $\mathcal{C}=\langle T(x_0)\rangle/\langle
  T(1)\rangle^2$ and $\lambda_0=\langle T(1)\rangle^{-1}$, as
  predicted by Eq.~(\ref{general}). Similarly, we find a perfect
  agreement
  between Eq.~(\ref{expo2}) and the known result
  for 2-dimensional diffusion (see e.g. \cite{GodecPRE}). Other cases can be checked analogously by explicit computation.

In contrast, for $0\le \nu\le \frac{3}{2}$, i.e. compact (recurrent) dynamics \cite{Liou}, where the particle
densely explores the space (e.g. 1-dimensional diffusion and diffusion on
fractals when $d_f<d_w$ \cite{oNChem}) the Newton
series in Eq.~(\ref{exact}) does not converge with the first term in the limit
$x_a\to 0$. Moreover, taking the limit $x_a\to 0$ we find that $\lambda_0$ becomes
independent of $x_a$ and hence the notion of a target size
effectively ceases to exist for $\nu<0$ such that
  taking the limit $x_a=\frac{r_a}{R}\to 0$ in this case corresponds to a change of
  the time unit, i.e. the result is invariant if we choose
  $\tau=(R-r_a)^{d_w}/D_{d_w}$.\\
\indent Conversely, for non-recurrent exploration ($\nu<0$), there is no such
invariance and for any choice of $\tau$, as the exponent $\lambda_0^{-1}$ contains
the term $x_a^{-|\nu|}$ (see Eq.~(\ref{expo})) such that
 $\lambda_0$ vanishes uniformly as
$x_a\to 0$. This demonstrates that this limit for non-compact
exploration corresponds mathematically to a genuine singular perturbation.
The limits $r_a\to 0$ at
fixed $R$ and $R\to\infty$ at fixed $r_a$, both giving $x_a\to
0$, are equivalent under an
appropriate change of time units and we already
demonstrated that there is no difference between compact and
non-compact FPT kinetics as long as $x_a\not\ll 1$. Taking the limit $x_a\to 0$
therefore has a very different mathematical meaning for $\nu\gtrless
0$. This is an important observation we will return to below.

\subsection{\emph{Deterministically biased diffusion}}

Turning now
from passive to biased (driven) diffusion, we first address
\noindent Taylor dispersion, the 1-dimensional
diffusion under the influence of a constant bias $v$ away from an
absorbing target at $0$, with a reflective boundary at $R$. 
Rescaling $x\to\frac{x}{R}$,  the
dimensionless P\'eclet number $\mathrm{Pe}=\frac{vR}{2D}$ captures all
dynamical regimes \cite{Redner}. The natural time unit is 
$\tau_T=\frac{D}{v^2}=\frac{R}{2v\mathrm{Pe}}$, and the dynamics is non-recurrent
(non-compact). The short time asymptotic reads (SI-S4) 
\begin{equation}
\label{sh_Taylor}
\wp(t)\sim
\Phi(t;x_0)\mathrm{e}^{-\mathrm{Pe}(x_0+\mathrm{Pe}t)}= (2\pi t^3)^{-1/2}\mathrm{e}^{-(x_0+2\mathrm{Pe}t)^2/4t}
\end{equation}
for $t\ll 1$, which apparently  deviates from the
Sparre Andersen form \cite{Sparre}. Instead, in the limit of a large bias towards the
target $\mathrm{Pe}\to -\infty$ the entire exponential term
effectively acts as a cut-off
at the deterministic FPT $t=x_0/2\mathrm{Pe}$. Note
that a detailed analysis of the record and first-passage statistics
of the corresponding discrete version of the model was studied in
Ref. \cite{SatyaD} using a random walk approach.

In the opposite limit when $t\to\infty$ 
and when $\mathrm{Pe}\gg 1$, the correction term given in Eq.~(S37) vanishes and
the Newton series in Eq.~(\ref{exact}) reduces to (SI-S4)
\begin{equation}
\label{expoT}
\lambda_0 \sim \mathrm{e}^{-2\mathrm{Pe}},\qquad \mathcal{C}\sim\mathrm{e}^{-2\mathrm{Pe}}(1-\mathrm{e}^{-2\mathrm{Pe}x_0}).
\end{equation}
This result has the universal form of Eq.~(\ref{general}).
When $\mathrm{Pe}\to-\infty$  we need an increasing number of terms in
Eqs.~(\ref{theorem}) and (\ref{exact}) as we approach the deterministic limit
 $\wp(t)\sim\delta(t-\frac{x_0}{2\mathrm{Pe}})$. This finding agrees with the
qualitative change of the behavior found in Ref.~\cite{SatyaD}, when the
external bias switches from strong repulsion to strong attraction.

Similar observations are made for $2$-dimensional diffusion under the influence of
a radial bias $v(r)=\frac{v_0}{r}$ \cite{Redner,Koplik} (SI-S5). We introduce 
$\mathrm{Pe}=\frac{v_0}{D}$ ($\mathrm{Pe}>0$ for outward bias; note
the different form with respect to the 1-dimensional case) and
dimensionless coordinates $x_{a,0}=\frac{r_{a,0}}{R}$. 
Due to the
analogy with passive diffusion in a fractal medium when
$\nu=-\frac{\mathrm{Pe}}{2}$ (see also \cite{Redner,Koplik}) we immediately read off the short-
and intermediate-time
asymptotics in Eqs.~(\ref{sh_fract}) and (\ref{int_fract}). For $\mathrm{Pe}\to \infty$, the long-time
behavior is given by \eqref{expo} and we recover the universal
form (\ref{general}).
As in the passive fractal case the limit $\nu\to 0$ for $x_a\ll 1$ in Eq.~(\ref{expo2})
is approached symmetrically for $\mathrm{Pe}\to 0$ from
below and above.
For $\mathrm{Pe}\to-\infty$ (inward bias) increasingly many
terms are needed in Eqs.~(\ref{theorem}) and (\ref{exact}) as we slowly approach the
deterministic limit
$\wp(t)\sim\delta\left(t-x_0^2/2|\mathrm{Pe}|\right)$ (SI-S5). 
 
As a second example we address the FPT problem for diffusion in a harmonic
potential, i.e. the Ornstein-Uhlenbeck (OU) process or a concrete
example for the Kramer's escape from a potential well \cite{Kramers,Siegert,Haenggi,Schulten,Eli}. Specifically, we
consider a particle with mass $m$ and friction coefficient $\kappa$
starting at $x_0$ and
diffusing in a harmonic potential $U(x)=m\omega^2x^2/2$ with an
absorbing boundary at $x_a$ (we choose
$x_0<x_a$) in the overdamped limit $\kappa\ll\omega$. We introduce
 the
characteristic length $l=\sqrt{k_BT/(m\omega^2)}$ of the potential gauged by the thermal energy
$k_BT$, rescale coordinates $x\to x/l$, and express time in natural units
$\tau_K=\kappa/(m\omega)^{2}$.
The OU process is
positive-recurrent \cite{Liou} and hence corresponds to compact
dynamics. When both $x_a$ and $x_0$ are not too large the short time
asymptotics are obtained using the expansions in \cite{AbSteg}. For
$t\ll x_a^{-2},x_0^{-2}$ (SI-S6), 
\begin{equation}
\label{sh_Kramers}
\wp(t)\sim
\mathrm{e}^{\frac{x_0^2-x_a^2}{4}}\Phi(t;x_a-x_0)\circ\sum_{i=1}^2Q_i\left[\frac{1}{\sqrt{\pi
t}}+\mathrm{e}^{P_i^2t}\mathrm{erfc}(P_i\sqrt{t})\right]
\end{equation}
 where $\circ$ denotes the one-sided
convolution, $f(t)\circ g(t)=\int_0^tf(t')g(t-t')dt'$. Here we use
\begin{eqnarray}
\label{coeffs}
Q_{1,2}&=&\frac{x_a^3+x_0^3}{48}\mp\frac{1}{\sqrt{1+144x_a^{-4}}}\left[\frac{3}{2}\left(\frac{1}{x_a}-\frac{x_0^2}{x_a^3}\right)+\frac{x_a^3+x_0^3}{48}\right],\nonumber\\
P_{1,2}&=&\frac{x_a^3}{48}\left(1\pm \sqrt{1+144x_a^{-4}}\right),
\end{eqnarray}
where the upper sign is associated with the index $1$. 
 The
correction terms, which are necessary when $x_a$ and $x_0$ become significantly
larger than $1$ (yet remain moderate), extend the range of validity to longer
$t$. In the case of $x_a,x_0\lesssim 1$, Eq.~(\ref{sh_Kramers})
reduces to 
\begin{equation}
\wp(t)\sim
\mathrm{e}^{(x_0^2-x_a^2)/4}\Phi(t;x_a-x_0),
\end{equation}
which apart from a $t$-independent prefactor is the Sparre
Andersen form \cite{Sparre}.
%Intermediate time-scale asymptotics, i.e. $t\gtrsim x_a^{-2},x_0^{-2}$, can be derived using Darwin's
%expansions \cite{AbSteg}. 

For the OU process the coefficients
$h^{(k)},g^{(k)}$ in Eqs.~(\ref{exact}) and (\ref{theorem}) are given in
terms of $H_0^{n,0}(x)\equiv(-1)^n\frac{\partial^n}{\partial
  \alpha^n}H_{\alpha}(x)|_{\alpha =0}$, where $H_{\alpha}(x)$
represent generalized Hermite functions, which are readily implemented,
for instance,
in Mathematica. For
large barrier heights the correction term in Eq.~(S47) vanishes and it can be shown (SI-S6) that the 
long-time FPT
asymptotic attains the universal form of Eq.~(\ref{general}) with
\begin{equation}
\label{Kramers}
\lambda_0 \sim \frac{x_a}{\sqrt{2\pi}}\mathrm{e}^{-x_a^2/2},\quad
\mathcal{C}\sim \frac{x_a}{\sqrt{2\pi}}\mathrm{e}^{-x_a^2/2}\left[1-\frac{x_a}{
\sqrt{2\pi}}\mathrm{e}^{-x_a^2/2}H_0^{1,0}\left(x_0\right)\right],
\end{equation}
and where we use the fact that
\begin{equation}
 H_0^{1,0}(x)=\frac{1}{2}\Bigg[\pi\mathrm{erfi}\left(\frac{-x}{\sqrt{2}}\right)-x^2{_2F_2}\left(\begin{matrix}
  1, & 1 \\
  3/2, & 2 
 \end{matrix};-\frac{x^2}{2}\right)-\delta\Bigg].
\end{equation}
Here $\delta=\gamma+2\log 2$ and $\gamma\simeq 0.5772$ is the Euler-Mascheroni
constant \cite{AbSteg} and we introduced the generalized hypergeometric function 
%${_2F_2}\left(\begin{matrix}  a_1, & a_2 \\  b_1, &
%  b_2 \end{matrix};z\right)$ 
${_2F_2}(z)$ \cite{AbSteg}. 
Note that since
$\lim_{x\to-\infty}H_0^{1,0}(x)\sim-\log(x)$, $\wp(t)$ only weakly
depends on $x_0$ for large initial separations. Conversely,
$\mathcal{C}$ can become very small in the case when $x_0\sim x_a$: Developing an asymptotic expansion via analytic continuation to
complex $x$ for Eq.~(\ref{Kramers}) and for the exact mean FPT
\begin{equation}
\label{MFPTKr}
\langle
T(x_0)\rangle=H_0^{1,0}\left(x_0\right)-H_0^{1,0}\left(x_a\right)
\end{equation}
 we recover the universal form in
Eq.~(\ref{general}). In the opposite case when $x_a\ll-1$ we approach
the deterministic limit $\wp(t)\sim\delta(t-\log\frac{x_a}{x_0})$ (see
SI-S6 and \cite{Siegert}). 
%The exact FPT asymptotic in
%\eqref{Kramers} and the exact $\langle T(x_0)\rangle$ for the OU
%process obtained here are our fourth main result.

\subsection{\emph{Discussion of the results in the context of
    molecular signaling in biological cells}}

\noindent Nature apparently developed at least two distinct strategies to
achieve both fast and precise molecular signaling in cells, necessary,
for instance, in gene regulation. One is the
energetically expensive and highly specific directed active transport
by hitchhiking molecular motors and
pertains to trafficking over large distances \cite{GodMetzRC,Alberts,OlivierAct,Granick,GodMetz}. The
second one is the spatial colocalization, i.e. proximity, of genes in
which the protein encoded by one gene regulates the second gene
\cite{otto,Kolesov,Fraser}. The kinetics of this
regulation process of the second gene include the diffusion of the
protein of the first gene to the second gene. 
%of successively activated
%genes \cite{otto,Kolesov,Fraser}, which according to our results is not
%limited to the geometry-controlled regime \cite{oNChem} but is of
%generic nature. 
Despite the variety of motion patterns (diffusion, subdiffusion, diffusion
on fractals etc.) observed in different experiments
\cite{Chro1,ChromRev,BialekBK,Gratton}, our results provide an
explanation of the robustness of the proximity
effect for efficient and precise transcription regulation.

Typically, in bacterial cells and eukaryotic nuclei, respectively, there are roughly $\lesssim 10^2$ \cite{Bialek,BialekBK} and
$\lesssim 10^4$ \cite{Zlat} transcription factors (TFs) of one kind searching for their
specific binding site on the cellular DNA. The linear extension of
bacteria and eukaryotic nuclei is $\sim 1\mathrm{\mu}$m
and $\lesssim 5\mathrm{\mu}$m, respectively
\cite{Bialek,Zlat,BialekBK,Eukaryot}. The vast majority of the DNA
binding sequences (promoters) 
is occupied by only a few proteins \cite{occup}. 
The promoter binding sites have a
typical size of $\sim 3$nm \cite{Bialek,Zlat,BialekBK,Eukaryot}, thus
$x_a\simeq 10^{-3}$. The typical distance between colocalized genes is of order $\simeq 30-300$nm (the size of a transcription
factory) \cite{Zlat,Fraser}, hence $|x_0-x_a|\simeq 5\times
10^{-2}$. 

\begin{figure}
\begin{center}
\includegraphics[width=16.cm]{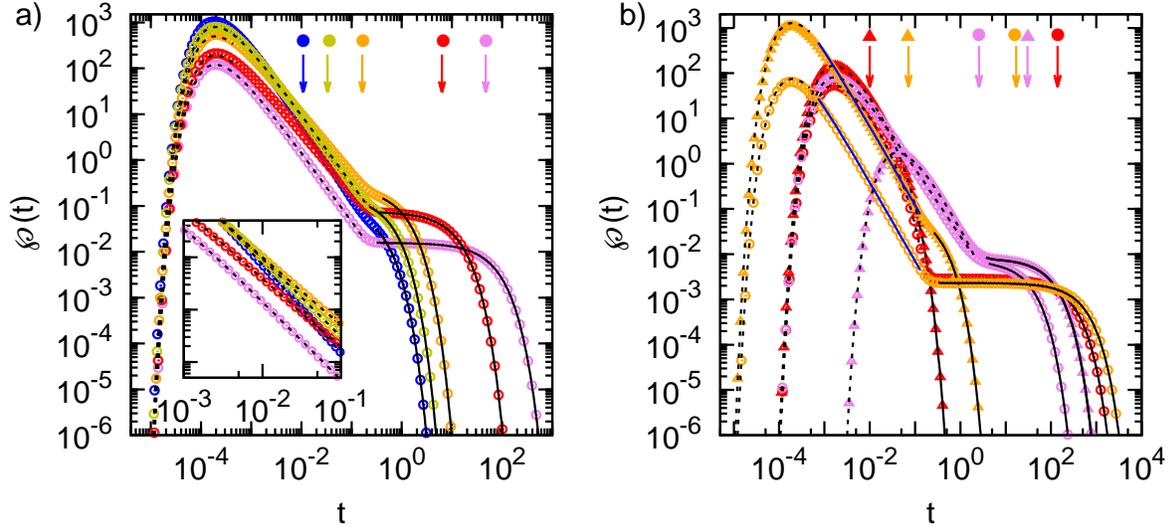}
\caption{FPT densities for various passive (a) and biased (b)
  diffusive systems along with mean FPT values (arrows). a) $x_a=0.006$, $x_0=0.04$ and $\nu=-1/2$ (violet), $\nu=-1/4$ (red), $\nu=-1/4$ (orange), $\nu=1/2$  (yellow)
and $\nu=0.7$-strong subdiffusion (blue). The symbols represent results
of numerical simulations (see Methods), full lines correspond to
\eqref{theorem} (or the Poissonian limit for $\nu<0$) and dashed lines to
\eqref{sh_fract}. Inset: Magnification of the intermediate regime
with the asymptotic (\ref{int_fract}) depicted by dashed
lines. Note that the long-time behavior for both $\nu<0$ cases is well
within the Poisson-like regime. b) 1d diffusion under constant bias for $x_0-x_a=0.1$ (orange; $\mathrm{Pe}=-5$
(circles) and $\mathrm{Pe}=5$ (triangles)), 2d diffusion with
radial bias for $x_a=0.006$, $x_0=0.04$ (red; $\mathrm{Pe}=0.75$
(circles) and $\mathrm{Pe}=-0.75$) and OU process (violet; $x_a=3$, $x_0=2.5$
(triangles), $x_a=2.5$, $x_0=2.4$ (circles)). The symbols represent results
of numerical simulations (see Methods), full lines correspond to
\eqref{theorem} (or the Poissonian limit for $\nu<0$), black dashed lines to
\eqref{sh_fract} and \eqref{sh_Kramers}, respectively, and the blue full lines to
\eqref{int_fract}.}
\label{Classes}
\end{center}
\end{figure}

Comparing $\wp(t)$ for different $\nu$ for these parameters on their respective
natural time-scale (Fig.~\ref{Classes}a) we observe identical qualitative behavior,
regardless of the compactness of the exploration. Namely, in all cases we
find a disparity of 2 to 4 orders of magnitude between the
likelihood for direct trajectories---corresponding to Eqs.~(\ref{sh_fract}) and (\ref{int_fract})---versus
indirect trajectories---Eqs.~(\ref{theorem}), (\ref{expo}) and  (\ref{expo2}). This peak of
accumulated probability mass is very narrow, i.e. 1 to 4 orders of
magnitude narrower than the respective time scale $\lambda_0^{-1}$. The
exponential long-time region is invariably statistically dominant
(somewhat less so for compact dynamics, where the MFPT is shifted into the
intermediate regime---compare the arrows in Fig.~\ref{Classes}a) but have a 
$\gtrsim 4$ orders of magnitude smaller value of $\wp(t)$. This means
that, while the majority of trajectories is indirect and falls into the
long-time regime, any two such indirect realizations will have
massively different FPT (see also \cite{Carlos2,Tiago}). Conversely, direct trajectories are focused, i.e. have
very similar FPT, giving rise to a sharp mode in the direct regime,
$t_{\mathrm{typ}}=(x_0-x_a)^2/6$. As it suffices that only a few of
the $10^2$-$10^4$ TFs ,i.e. the fastest 0.01\% to 1\%, actually need to bind their target, the numerous, strongly dissimilar,
indirect realizations are in fact
irrelevant. Gene regulation kinetics as uncovered by
  modern experimental techniques in single cells \cite{Stoch,xxie,cai,honigman,weigel1} therefore occur in
the \emph{few-encounters limit}, for which our results are
relevant. Conversely, the results of Ref. \cite{oNChem} are
relevant if we ask for the typical FPT of \emph{any} of the TF to
their site, as would be measured in a bulk experiment for the behavior of an
ensemble of genetically identical cells in a colony.
In this sense the results found here readily
explain both the speed- as well as the precision-enhancement for the
regulation of colocalized genes, irrespectively of the compactness of
spatial exploration.
The proof and explanation of the
proximity effect in transcription regulation is our second main result.
These principles can readily be extended
to the more complex facilitated diffusion model
\cite{otto,Kolesov,BergHipp}. 

Further substantiating the robustness of the proximity effect is the fact that qualitatively the results remain unchanged in the presence of a bias,
Fig.~\ref{Classes}b. Naturally, a bias towards the target shifts the
long-time regime towards shorter $t$. What might appear somewhat
surprising at first, is that it does not affect, apart from a trivial renormalization, the
statistics, i.e. the width and position of direct trajectories
(Fig.~\ref{Classes}b). This observation is, however, a
straight-forward consequence of the independence of drift and
diffusion. It highlights the intrinsic universality of direct paths
and conceptually extends the Sparre Andersen universality
\cite{GodSciRep}. Remarkably, proximity effects persist in the OU
process (Fig.~\ref{Classes}b-violet), for which direct, strictly uphill
trajectories are typical and in turn imply excessive fluctuations of
FPT times of indirect trajectories.

\section{Concluding remarks}

Our rigorous results provide a general
  method to determine the exact long-time behavior of a smooth
one-sided probability density with finite moments from its Laplace
  transform. They extend the existing generic asymptotic inversion methods,
  such as Tauberian theorems for slowly varying functions, to
  functions varying exponentially fast at long times.
  Here this new method enabled us to unify the first passage time
  statistics of all FPT densities with asymptotic exponential
  decay, both passive as well as those under the influence of a
  deterministic bias, into a single universality class. Moreover, they provided a
  deep physical insight into the general effects of (non)recurrent
  spatial exploration on the FPT behavior, and highlighted the
  qualitatively very different meaning of assuming a large volume or a
  small target size. Beyond the theoretical advance our results are
  directly relevant in a biophysical context.

Thus, modern experimental methods allow the observation of individual
molecular regulation events. It is therefore timely to extend the classical
vista of mean rate approaches to biochemical kinetics and consider the
full distribution of FPTs  \cite{oberg,carmine}.  The theoretical results
presented here will be a quantitative basis for the development and analysis
of massive single-molecule  experiments for FPT dynamics in living cells
or chemical reactions in micro- and nano-containers. The few-encounter limit
introduced here together with the proximity effect complements the many-encounter
regime associated with the mean FPT theory developed in Ref. \cite{oNChem}.
In this latter regimes we fully corroborate the results of Ref. \cite{oNChem}, in
particular, the observation that the initial distance to the target will be
critically more important for compact exploration.

Our results have immediate consequences for the interpretation of single-molecule
reaction or binding experiments (e.g. \cite{Kurzinsky} and refs. therein), that
nowadays allow one to directly visualize the few-encounter regime. Namely,
the quantification of reaction kinetics is paradigmatically
reduced to the mean FPT \cite{Haenggi,Szabo,Katja}, which is
meaningful if we are interested in
systems, in which most of the molecules are required to react. In contrast, when only a small number of
reactive encounters are required, e.g. the binding of
a few TFs to their targets deciding
the regulatory pathway of a cell \cite{Stoch} or phosphorylation cascades in cellular signal-transduction
propagating towards the nucleus \cite{Wolynes}, our results show that typical
and direct realizations are essential. In this few-encounter regime, mean FPT-based
concepts grossly underestimate the speed and precision of
experimentally observed signaling kinetics, and would thus lead to severe
parameter misjudgments. These would deteriorate further for FPT cascades.   
In other words, the notion of a kinetic rate in the
traditional bulk sense ceases to exist in the few-encounters regime. 
The universality of the proximity effect in target search kinetics enabling temporal signal focusing
therefore challenges traditional views on biochemical reactions in cells 
and provides the basis for new models for molecular regulatory kinetics in
the few-encounters regime.

\begin{acknowledgments} We thank Olivier B\'enichou for instigating the present work
  and for discussions. AG acknowledges funding through an Alexander von Humboldt
Fellowship and ARRS project Z1-7296.
\end{acknowledgments}

\end{document}